\documentclass[aps,prd,preprint,nofootinbib]{revtex4-1}
\usepackage{amsmath,amsfonts,amssymb,hyperref}
\usepackage[english]{babel}
\usepackage[utf8]{inputenc}
\usepackage{amsthm}
\usepackage{balance}
\usepackage{mathtools}
\usepackage{physics}
\usepackage{xcolor}
\usepackage{graphicx}
\usepackage[left=23mm,right=13mm,top=35mm,columnsep=15pt]{geometry} 
\usepackage{adjustbox}
\usepackage{placeins}
\usepackage[T1]{fontenc}
\usepackage{lipsum}
\usepackage{csquotes}
\usepackage{verbatim}

\def\be{\begin{equation}}
\def\ee{\end{equation}}
\def\ber{\begin{eqnarray}}
\def\eer{\end{eqnarray}}
\def\bwt{\begin{widetext}}
\def\ewt{\end{widetext}}

\begin{document}
\title{
Recent Efforts Towards Understanding the Early Universe from a Fundamental Quantum Perspective}

\author{Sujoy K. Modak}    
    \email{smodak@ucol.mx}
    \affiliation{Facultad de Ciencias - CUICBAS, Universidad de Colima, Colima, C.P. 28045, M\'exico}
    \affiliation{Instituto de Ciencias Nucleares, Universidad Nacional Aut\'onoma de M\'exico, POB 70-543, 04510, CDMX, M\'exico}

\date{\today} 

\begin{abstract}
The observable universe is fundamentally \emph{inhomogeneous} and \emph{anisotropic}. Quantum description of the generation of these inhomogeneities and anisotropies is  ill-understood and unsatisfactory. After providing a brief account of the standard approach of the generation of the classical density perturbations starting from the quantum fluctuations of inflaton field, I critically review various assumptions which are crucial for the success of this description, and point out various shortcomings around it. I also discuss the basic ideas and recent works by using an alternative path to overcome those shortcomings which is motivated by the so-called Collapse Model interpretation of quantum mechanics. Inspired by these works, I argue the necessity of constructing a class of manifestly inhomogeneous and anisotropic quantum states after inflation and discuss my recent works which provide one such prescription of building such a state, called the $T-$vacuum, defined in the radiation dominated stage of the early universe. 



\end{abstract}


\maketitle

\tableofcontents

\section{Introduction}
For long our central source of information about the structure and the evolution of the universe were restricted to the observations of redshifts from distant galaxies at the astrophysical scale. In 1965 a nearly isotropic background of microwave radiation was discovered which revolutionized our understanding of the universe due to an access to a large cosmological data. Observations on the distant galaxies provided us an information, even before the observed microwave radiation, that the universe is in an expanding phase. If we have to believe that the expansion of the universe is a feature from the earliest of times, then we could imagine that in the past the matter was hotter and denser than present. In a distant past the temperature was so high that electrons were not bound with atoms and radiation in the form of photons were in collisions with matter and hence the radiation was in thermal equilibrium with matter. The number density of photons in equilibrium with matter at temperature $T$ followed a black-body distribution. As time passed matter become cooler and denser forming nucleus and then atoms but the photons got separated and began a free expansion, but its \emph{spectrum kept the same form}. Usually, it is assumed that there was a time $t=t_L$ where the radiation got suddenly separated from matter, and this $L$ stands for the surface of the ``last scattering''. For the expanding universe with scale factor $a(t)$ the frequency $\nu$ of freely moving photons at a time $t>t_L$ would have had a frequency $\nu a(t)/a(t_L)$ at the time of decoupling. The effect of cosmic expansion is such that the form of the blackbody spectrum remain unaffected during the free expansion after photons went out of equilibrium with matter just the temperature gets a modification. This modification relates the temperature at any later time after this ``photon decoupling'' with the temperature at the last scattering, given by $T(t)=T(t_L)a(t_L)/a(t)$. While the instantaneous decoupling of photons is an approximation, it is quite a convincing one as frequent elastic and infrequent inelastic collisions even during the transparent photon era would not modify the blackbody spectrum by much. 

George Gamow was one of the first to realize the possible existence such a background spectrum \cite{gam} but it was not until 1965 when a detection was made by the radio telescope by Penzias and Wilson \cite{pen} which could only detect the radio wavelength of 7.5 cm and a temperature of about 3.5 $\pm$ 1.0 K. 

Although this Cosmic Microwave Background (CMB) radiation is highly isotropic there exist several types of relatively small anisotropies and these characteristics of CMB provide some of the most revolutionary information about our universe. The frame of CMB provides a natural frame of reference for studying the universe.  Among the various types of anisotropies, that do appear in the CMB temperature distribution over the sky, one is due to the Earth's own motion, and the temperature of CMB varies depending on which direction we look up to. Earth's own motion with respect to the CMB frame depends on the motion of the solar system as well as the rotation of the galaxy itself. According to the WMAP satellite experiment the net velocity of the local group of galaxies relative to the microwave background is 627$\pm$ 22 km/s in a specific direction given by an azimuthal angle $\ell = 276 \pm 3$ degrees and the  angle between the galactic plane with the line of sight $b = 30 \pm 3$ degrees \cite{wein}. The second type of anisotropy is due to the Sunyaev-Zel'dovich effect \cite{SZ} which attributes to the scattering of radiation by electrons in the intergalactic space in between the clusters of galaxies along the line of sight. Apart from these secondary anisotropies which are caused by the late universe and appear during the journey of CMB photons to their way to us, there is also primary type of anisotropies that have their origin in the early universe \cite{reese}. The primary anisotropies are originated at the Last Scattering Surface (LSS at $z=1090$) due to various processes, for example, (i) an intrinsic temperature fluctuations in the electron–nucleon–photon at the time of last scattering, (ii) the Doppler effect due to velocity fluctuation in plasma, and (iii) Sachs-Wolfe effect which is redshift/blueshift due to a fluctuating gravitational potential \cite{SW}. 

In this article I review the physics behind the origin of the tiny anisotropies at the LSS from a quantum standpoint. The reason behind this is the fact that the in the inflationary universe, and even a short (but unknown) period after inflation, all matter fields were quantum, and the fluctuations of these quantum fields, famously named ``inflaton'', are  the source of these observed angular temperature fluctuations in the CMB. This article is organized as follows: in the next section \ref{sec2} I discuss the empirical formula that can be constructed using the temperature fluctuations and relating those with the fluctuations of the classical density perturbations. We introduce the famous Harrison-Zel'dovich power spectrum which is what is observed from CMB.  Next, in section \ref{sec3} I provide an elaborate discussion, based on the monograph by Weinberg \cite{wein}, describing standard quantum description that is most famous in the literature. Here, I will be using a critical stance, strictly  reviewing the underlying assumptions, and in the next section \ref{sec4} I will refute all four popular assumptions, which will lead us to the foundation problems with quantum theory. In section \ref{sec5}, I shall present new alternative path to understand the quantum description using the so-called Collapse Model interpretations of quantum theory. Finally, in section \ref{sec6} I will present some new results derived by us in order to understand better quantum fields in radiation dominated universe. This last section is a complementary path to define a set of physically motivated anisotropic quantum vacuum states with an ambition that finally we may be able to point these works down to yet unknown {\emph{quantum state}} of the early universe at LSS (or before) generating the density perturbations more convincingly.

\section{CMB temperature fluctuation and the power spectrum}\label{sec2}

In the integrated Sachs-Wolfe effect the perturbation to the gravitational potential is a time dependent function $\delta \phi(\bf{x})$ where $\bf{x}$ is the comoving coordinate. This perturbation slightly changes the frequency of the emitted photon from the point $\bf{x}$ on the LSS, and its energy is also slightly shifted. This implies that the temperature obtained by looking in a particular direction $\hat{n}$ is shifted from the averaged value over the whole sky, given by
\begin{equation}\label{fs1}
    \frac{\Delta T(\hat{n})}{T_0} = \delta\phi (\hat{n} R_{LSS}),
\end{equation}
where \cite{wein}
\begin{widetext}
\begin{equation}
    R_{LSS} = \frac{1}{H_0 a(t_0) \sqrt{\Omega_K}} \sinh\left[ \sqrt{\Omega_K} \int_{1/(1+z_L)}^{1} \frac{dx}{\sqrt{\Omega_{\Lambda} x^4 + \Omega_{K} x^2 + \Omega_M x + \Omega_R}}\right]
\end{equation}
\end{widetext}
is the radial coordinates for the LSS, $\Omega_\lambda, \Omega_M, \Omega_R$ are the fractions of energy densities corresponding to vacuum, non-realtivistic and relativistic matters and $\Omega_K = 1-\Omega_M - \Omega_R - \Omega_K$, $K=0,\pm 1$ (spatially flat, open/close), $x=a/a_0=1/(1+z)$ ($z$ is the redshift factor). The subscript zero belongs to the present value and $z_L=1090$ is the redshift of the LSS. The perturbation to the gravitational potential also affect the rate of expansion of the universe that also leads to a fluctuation of the temperature given by 
\begin{equation}\label{fs2}
    \frac{\Delta T(\hat{n})}{T_0} = -\frac{2}{3}\delta\phi (\hat{n} R_{LSS}).
\end{equation}
The total fractional change in temperature is then the sum of \eqref{fs1} and \eqref{fs2}, and is given by
\begin{equation}\label{fs2}
    \frac{\Delta T(\hat{n})}{T_0} = \frac{1}{3}\delta\phi (\hat{n} R_{LSS}).
\end{equation}
which is known as the Sachs-Wolfe effect which is related with the perturbation $\delta \rho$ in the total mass density through the Poisson equation, given by
\begin{equation}
    \frac{1}{a^2} \nabla^2\delta\phi ({\bf{x}})= 4\pi G \delta\rho ({\bf{x}}).
\end{equation}
This equation, after expressing the Fourier transformation of $\delta\rho ({\bf{x}})$ in terms of $\delta\phi (\bf{x})$ yields the correlation function of the density fluctuations
\begin{widetext}
\begin{equation}
    \langle \delta\rho ({\bf{x}},t) \delta\rho ({\bf{x}}',t') \rangle = \frac{1}{(4\pi G a(t) a(t'))^2} \int {d^3k~ ~{\cal P}_{\phi}(k) e^{i\bf{k}.({\bf{x}} - {\bf{x}}')}},
\end{equation}
\end{widetext}
where ${\cal P}_{\phi}(k)$ is the power spectrum which is conventionally expressed as
\begin{equation}
    {\cal P}_{\phi}(k) = N_{\phi}^2 k^{n}.\label{hzs}
\end{equation}
Observations of the density correlation function  has shown that the above expression takes the so-called Harrison-Zel'dovich form  \cite{HZ} with $N_{\phi} \simeq 10^{-5}$ and $n=1$.

\section{The standard quantum description of the power spectrum}\label{sec3}
One of the celebrated successes of inflationary cosmological theories lies in the fact that it provides a natural quantum mechanical origin of cosmological fluctuations observed in CMB as described in the last section and in the large scale structure of matter. In this section I shall briefly discuss the standard picture keeping a close attention to various assumptions and hypothesis that are at play.    

I present here only the relevant part for this article following  Weinberg's monograph \cite{wein}. My aim is not repeat everything that is already discussed there but to give a much condensed version with a special focus on the basic structure beneath the complete construction.

Let us consider the simplest model of inflation with a single inflaton field $\phi (x)$, given by the action
\begin{equation}
    I = \int d^4x \sqrt{-g} \left(\frac{1}{2} g^{\mu\nu} \partial_\mu\varphi \partial_\nu \varphi - V(\varphi)\right),
\end{equation}
where $V(\varphi)$ is an arbitrary real potential. The first step is to express the scalar field as a sum of unperturbed $\bar\varphi (t)$ plus perturbation $\delta\varphi ({\bf{x}},t)$ that depends on both space and time, given by
\begin{equation}
    \varphi({\bf{x}},t) = \bar\varphi (t) + \delta\varphi ({\bf{x}},t).
\end{equation}
Similarly, the metric is given by the unperturbed Friedman-Robertson-Walker metric $\bar{g}_{\mu\nu} (t)$ plus a small perturbation $h_{\mu\nu} ({\bf{x}},t)$,
\begin{equation}
    g_{\mu\nu} ({\bf{x}},t) = \bar{g}_{\mu\nu} (t) + h_{\mu\nu} ({\bf{x}},t).
\end{equation}

The unperturbed Friedman equation (with $K=0$) is
\begin{equation}
    H^2 = \frac{8\pi G}{3}\left(\frac{1}{2}\dot{\bar\varphi}^2 + V(\bar{\varphi}) \right)
\end{equation}
while the  field equation for the unperturbed inflaton field is
\begin{equation}
    \Ddot{\bar{\varphi}} + 3H \dot{\bar{\varphi}} + V'(\bar{\varphi}) =0
\end{equation}

Now for the perturbations one may consider the Newtonian gauge to select $h_{00} = -2\Psi, ~h_{0i} = 0,~ h_{ij} = -2a^2\delta_{ij}\Psi$. The Einstein equation and energy conservation equations for the perturbations are 
\begin{widetext}
\begin{eqnarray}
   && \dot{\Psi} +H \Psi = 4\pi G \dot{\bar{\varphi}}\delta\varphi \label{per1}\\
   && \delta\Ddot{\varphi} + 3H\delta\dot{\varphi} + \partial^2_{\bar{\varphi}}V(\bar{\varphi})\delta\varphi - \frac{1}{a^2}\nabla^2\delta\varphi = - 2\Psi {\partial_{\bar{\var\phi}}}V(\bar{\varphi}) + 4\dot{\Psi}\dot{\bar{\varphi}}\nonumber\\ \label{per2}
\end{eqnarray}
while the constraint equation can be simplified using the relationship $\dot{H}=-4\pi G \dot{\bar\varphi}^2$ and is given by
\begin{equation}
    \left(\dot{H} - \frac{1}{a^2}\nabla^2\right) \Psi = 4\pi G \left(-\dot{\bar\varphi}\delta\dot{\varphi} + \Ddot{\bar{\varphi}}\delta\varphi\right). \label{per3}
\end{equation}
\end{widetext}

Once the above set of differential equations \eqref{per1}-\eqref{per3} describing the perturbations of the metric and the inflaton field are obtained the next step is to find the complete set of field modes which can be used to express the fluctuations as
\begin{eqnarray}
    \delta\phi ({\bf x},t) = \int d^3k \left[\delta\varphi_k(t) e^{i{\bf k}.\bf{x}} a_k + \delta\varphi^*_k(t) e^{-i{\bf k}.\bf{x}} a_k^* \right],\label{dphi}\\
    \Psi ({\bf x},t) = \int d^3k \left[\Psi_k(t) e^{i{\bf k}.\bf{x}} a_k + \Psi^*_k(t) e^{-i{\bf k}.\bf{x}} a_k^* \right].\label{psi}
\end{eqnarray}

At sufficiently early times one has $k/a >> H$ and $k/a >> \partial_{\bar\varphi}^2 V(\bar{\varphi})$, and the solution of the field equations \eqref{per1}-\eqref{per3} in their mode form satisfy the initial conditions as $a(t)\rightarrow 0$,
\begin{eqnarray}
    \lim_{a(t)\rightarrow 0} \delta\varphi_k(t) = \frac{1}{(2\pi)^{3/2}a(t)\sqrt{2k}} e^{-ik\int_{t_*}^{t}dt'/a(t')}, \label{inc1}\\
    \lim_{a(t)\rightarrow 0} \Psi_k(t) =  \frac{4i\pi G \dot{\bar{\varphi}}(t)}{(2\pi)^{3/2}a(t)\sqrt{2k}} e^{-ik\int_{t_*}^{t}dt'/a(t')} \label{inc2},
\end{eqnarray}
while their complex conjugates are the other set of independent solutions. Also, in the early times the coefficients $a_k$ and $a_k^*$ can be identified with the creation and annihilation operators. Note that, although here the same creation and annihilation operators appear in the scalar field and gravitational perturbations $\Psi$ does not correspond to gravitational radiation, rather it is an auxiliary field and a functional of the inflaton field $\delta\varphi$ in the similar way that in the Coulomb gauge quantization of quantum electrodynamics the time-dependent component of the vector potential is a functional of the charged matter fields. The vacuum expectation values of the paired fields are given by
\begin{eqnarray}
    \langle 0| \delta\varphi ({\bf{x}},t) \delta\varphi({\bf{y}},t)|0\rangle = \int d^3k |\delta\phi_k(t)|^2 e^{i{\bf{k}}.({\bf{x}} - {\bf{y}})}, \label{tp1}\\
    \langle 0| \Psi ({\bf{x}},t) \Psi({\bf{y}},t)|0\rangle = \int d^3k |\Psi_k(t)|^2 e^{i{\bf{k}}.({\bf{x}} - {\bf{y}})}, \label{tp2}\\
    \langle 0| \delta\varphi ({\bf{x}},t) \Psi({\bf{y}},t)|0\rangle = \int d^3k \delta\phi_k(t) \Psi^*_k(t) e^{i{\bf{k}},({\bf{x}} - {\bf{y}})}, \label{tp3}\\
    \langle 0| \Psi ({\bf{x}},t) \delta\varphi({\bf{y}},t)|0\rangle = \int d^3k \Psi_k(t) \delta\phi^*_k(t) e^{i{\bf{k}}.({\bf{x}} - {\bf{y}})}. \label{tp4}
\end{eqnarray}
As mentioned in the literature \cite{wein,lyth} that clearly these are quantum averages and not the averages over an ensemble of classical field configurations which is evident from equations \eqref{tp3} and \eqref{tp4}. These two equations produce complex results for the averages of products of real scalar fields. However, it is critical that they reproduce classical results of density perturbations that will eventually be seen in the CMB map. In order that to happen these averages, although quantum and complex in the beginning of inflation, must lead to classical and real values at the end of inflation for the appropriate field modes which contribute to the primordial inhomogeneities in the classical density and an anisotropy over the CMB sky. This is a challenging task for various reasons one of them is of course to explain satisfactorily how this quantum-to-classical transition might have taken place.

There are three assumptions are made to explain the emergence of classicality:
\begin{itemize}
    \item \underline{Assumption 1}: Just as in the measurement of the spin in the laboratory a decoherence will take place making the above-mentioned quantum averages \eqref{tp1}-\eqref{tp4} ``classical''. 
    \item \underline{Assumption 2}: This quantum-to-classical transition happen when the perturbations exit the Hubble horizon. These perturbations $\delta\phi_k$ and $\Psi_k$ become classical and are locked into one of the ensembles of classical configurations. Hence can be treated classically.

    \item \underline{Assumption 3}: Once the universe become ``classical'' in the above sense one may use the so-called Ergodic Theorem \cite{wein} to interpret averages over ensembles of possible classical universes as averages over the position of the observer in \emph{our universe}. 
\end{itemize}
The first goal of this article is to present some concrete counter-arguments of all the above assumptions that have been made in various recent works and to show that none of the assumptions are satisfactory enough to resolve the issue at hand. In fact, these assumptions are inevitably related with the foundation of quantum theory -- the so-called ``Measurement Problem'' -- which till date is unresolved.

But before we delve into such arguments let us first complete the review of the remaining steps leading us to the standard explanation of the Harrison-Zel'dovich power spectrum for the CMB anisotropies.

Since the observational quantities related with the cosmological fluctuations are outside the horizon, after inflation it is not necessary to calculate the full set of equations \eqref{per1}-\eqref{per3} with the initial conditions \eqref{inc1} and \eqref{inc2} for which one also has to select a potential $V(\varphi)$. Rather one focuses on a quantity ${\cal R}_k = -\Psi_k + H \delta\varphi/\dot{\bar{\varphi}} $ which is conserved outside the horizon during inflation. The aim is to use this quantity to provide an initial condition for the evolution of perturbations after they re-enter the horizon later but before than the decoupling time. The function ${\cal R} ({\bf x},t)$ is related with ${\cal R}_k$ much to the same way as perturbations in equations \eqref{dphi} and \eqref{psi} while the quantum average is given by
\begin{equation}
    \langle 0|{\cal R} ({\bf x},t) {\cal R} ({\bf y},t)|0 \rangle = \int d^3k ~|R_k(t)|^2~ e^{i{\bf{k}}.({\bf{x}} - {\bf{y}})}.\label{tpr}
\end{equation}
It is easier to solve ${\cal R}_k(t)$ directly than solving for $\delta\varphi_k(t)$ or $\Psi_k(t)$, however, usually a different gauge (than Newtonian) is chosen for a simpler calculation. The vacuum is defined in the early times as $a(t)\rightarrow 0$, for the modes satisfying $k/a>> H$ which are essentially same as free massless real scalar field modes in Minkowski spacetime, and it is simply given by $a_k|0\rangle = 0$. The assumption is that the state of the universe during inflation is the vacuum $a_k|0\rangle = 0$ which is known as the Bunch-Davies vacuum \cite{BD}. This is the fourth assumption:
\begin{itemize}
    \item \underline{Assumption 4}: the state of the universe during inflation is in the Bunch-Davies vacuum $a_k|0\rangle = 0$, in \eqref{tp1}-\eqref{tp4} and \eqref{tpr}, which is defined in the early times of inflationary epoch as $a(t)\rightarrow 0$.
\end{itemize}

This Fourier transform of ${\cal R} ({\bf x},t)$ satisfy the Mukhanov-Sasaki equation \cite{MS}
\begin{equation}
    \frac{d^2 {\cal R}_q}{d\eta^2} + \frac{2}{\eta}\frac{d{\cal R}_q}{d\eta} + q^2 {\cal R}_q = 0, \label{mseq}
\end{equation}
where $\eta=\int \frac{dt}{a(t)}$ is the conformal time. The initial condition, for $a(t)\rightarrow 0$, is given by
\begin{equation}
    {\cal R}_k(t) = -\frac{H(t)}{(2\pi)^{3/2}\sqrt{2k}a(t)\dot{\bar{\varphi}}} e^{-ik\int \frac{dt'}{a(t')}}.\label{inc3}
\end{equation}
While integrating \eqref{mseq} one considers the limit from $a=0$ to beyond the horizon for which $q/a<<H$. In this limit it is possible to drop the last term \eqref{mseq} multiplying $q^2$ and then there are basically two solutions for ${\cal R}_k$. One of those is decaying while the other is a nonzero constant ${\cal R}_k={\cal R}_k^{0}$. This constant value ${\cal R}_k^{0}$ in the super-Horizon limit is used for the analysis of cosmological fluctuations.

The quantity ${\cal R}_k^{0}$ is independent of the nature of inflaton potential in the deep sub-Hubble region while it is constant in the deep super-Hubble scale. However, the potential $V(\varphi)$ plays its role while the perturbations becomes super-Hubble from the sub-Hubble scale as $a(t)$ increases with time -- a phenomena usually named as the ``Hubble exit'' of the perturbations. This feature of the inflationary era, while imprinted on the cosmological perturbations, can be revealed by observing the scalar fluctuations in the later stages of the expanding universe.

This property of Hubble exit in the inflationary universe provide a bound on the maximum number of e-foldings during the inflationary period before the beginning of the next epoch (the radiation dominate era) which, for the slow-roll inflation, is given by
\begin{equation}
    {\cal N}_0 = \ln\left(\frac{\rho_{r}^0}{0.037~h ~\text{eV}} \right),
\end{equation}
where $\rho_{r}^0$ energy density at the beginning of the radiation dominated era. Putting the values of  other quantities one can calculate ${\cal N}_0 \simeq 68$ which means that we are restricted to explore these 68 e-foldings of inflation by studying the scalar perturbations. 

As mentioned above, the particular value of the scalar perturbation ${\cal R}_k^{0}$ does not depend on the particular choice of the potential but on the slow-roll character of the same, and this is helpful to carry out a specific calculation with a potential that obey such a property, for example $V(\varphi) = A e^{-\lambda \varphi}$ where $A,~\lambda$ are constants. The final result for ${\cal R}_k^{0}$ in the limit $q/a<<H$ can be expressed as \cite{wein}
\begin{equation}
    {\cal R}_k^{0} = -i\frac{\sqrt{16\pi G}}{k^{3/2}} \left(\frac{H(t_k)}{8\pi^{3/2}\sqrt{\epsilon}} \right),
\end{equation}
where $t_k$ is the comoving time for Hubble exit. The power spectral function for this solution is then given by the factor $k^4|{\cal R}_k^{0}|^2$, which is proportional to $k$, and therefore matches with the Harrison-Zel'dovich spectrum as discussed in the previous section and given by equation \eqref{hzs}. The theoretical restriction on the value of the constant ${\epsilon}<1$ and it is defined as $\epsilon = -\frac{\dot{H}}{H^2}=\frac{\lambda^2}{16\pi G}$. Observations, such as WMAP, is able to fix the value of this parameter which is found to be (third year WMAP result) $\epsilon = 0.021\pm 0.008$ which is in accordance with the theoretical restriction. 

This completes our brief review of the standard approach to describe the observed anisotropies based on the cosmological perturbation theory using inflaton field with a general but slowly rolling potential. While it is remarkable that the correct form of the Harrison-Zel'dovich spectrum \eqref{hzs} is obtained it is also important to review necessary assumptions that are at the core of this approach which we want to perform in the next section.

\section{Unsatisfactory issues with the standard approach}\label{sec4}
In this section I review the four assumptions that are listed in the previous section and discuss if the arguments behind these assumptions are acceptable beyond criticisms and why there is a wide range of disagreement on their acceptability. 
\begin{itemize}
    \item {\underline{Reviewing assumption 1}}: in some important works Perez, Sahlman and Sudarsky \cite{Perez:2005gh,dan2} had pointed out several shortcomings that are associated with the claim of quantum origin of the classical density perturbations. One cannot compare the appearance of classical result starting from \eqref{tp1}-\eqref{tp4} to a laboratory situation of spin measurement. While in the latter there is an ``observer'' performing the act of ``measurement'' by an appropriate use of an ``apparatus'', {\emph none} of these is true for the situation at hand in the early universe where we are describing the vary fact of structure formation. No physical entity can perform any such act of measurement and therefore assumption 1 described in the previous section cannot be objectively acceptable. As Weinberg mentions in his monograph \cite{wein} {\emph{``... decoherence cannot occur until expectation value of products of real fields become real...''}}, and therefore decoherence do not make the complex quantum averages into real statistical averages by itself. The unitary property of quantum theory does not allow this to happen and we do need something extra such as a spontaneous collapse of the wavefunction which is the topic for the next section. 

     \item {\underline{Reviewing assumption 2}}: although it is often argued that  when the perturbations exit the
Hubble horizon during the inflationary period they suddenly become classical and get locked in that classical state until they reappear inside the Hubble horizon during the radiation dominated era and detected in the LSS. In an important work by Singh, Modak and Padmanabhan \cite{Singh:2013bsa} a systematic study was performed to review this particular assumption. We used a Schrodinger picture and studied the evolution of real, massive scalar field in a toy universe with three stages of expansions given by the inflationary stage, radiation dominated stage and late time de Sitter stage. Using a notion of the strength of ``classicality'' as a measure of standard correlation between the field modes and its momentum via Wigner function, developed in some earlier works \cite{Mahajan:2007qc, Mahajan:2007qg}, we could quantify the level classical behavior for such modes. This model although is not exact to the quantization of both the gravitational and inflaton perturbations at the same time, it is in fact a very good approximation and in line with a semiclassical viewpoint where one only quantize the fields but not the background metric. Our result showed interesting properties -- (i) we could show that the degree of classicality is maximum when the field modes exit the Hubble radius during the inflationary period, which is in agreement with the first part of the assumption made in \cite{wein}, however, we also found (ii) that the second part of the assumption is violated since the degree of classicality oscillates when the same mode {\emph{reenters the Hubble scale during the radiation dominated era}}. Therefore, they do not get locked in any classical configuration and the hypothesis that those modes are responsible for  the classical density perturbations in the LSS is misleading.  

      \item {\underline{Reviewing assumption 3}}: the validity of this assumption needs the validity of the first two assumptions which, as we have discussed above, are not acceptable from a more fundamental point of view. We may need new physics in order to establish the first two hypotheses and once this objective is ensured then one may review the ``Ergodic Theorem'' as described by Weinberg \cite{wein}.

       \item {\underline{Reviewing assumption 4}}: the assumption that the quantum state of the universe during inflation is given by the Bunch-Davies vacuum has a serious problem which was also pointed out by Perez, Sahlman and Sudarsky \cite{Perez:2005gh, dan2}. The main objection is the following -- one cannot explain the observed anisotropies after inflation starting from quantum averages \eqref{tp1}-\eqref{tp4} or \eqref{tpr} where the quantum state $|0\rangle = |0\rangle_{\text{BD}}$, the initial conditions are \eqref{inc1}, \eqref{inc2} and \eqref{inc3}, while the evolution during the inflationary period is all \emph{unitary}. This is simply because the Bunch-Davies state, by definition is homogeneous and isotropic, and no unitary evolution can break this symmetry and is able to dynamically produce anisotropies which we eventually observed in CMB and associate with the density perturbations on the LSS. 
\end{itemize}

We therefore conclude that none of the assumptions mentioned in the section \ref{sec3} and used in the literature (as summarized in \cite{wein}) {\emph{ are not}} convincing enough and there is a scope of new ideas to eventually reproduce the Harrison-Zel'dovich spectrum \eqref{hzs} from other viewpoints.

\section{Alternative proposals using the objective Collapse Models}\label{sec5}


The quest for satisfactory resolution of the ``Measurement Problem'' gave birth to several  versions of quantum mechanics, and among them are (a) Many World Interpretation (MWI) \cite{mwi}, (b) Bohmian Mechanics \cite{BM} and (iii) Objective Collapse Models \cite{col-rev-1}-\cite{Bassi:2012bg}. Considerations for addressing the generation of classical density perturbations is a common topic that has been addressed from all of the above three versions except for the  MWI, for which this issue the problem does not need any new explanation. However, using the Bohmian approach this problem was addressed to some extent \cite{Goldstein:2015mha, Pinto-Neto:2018zvn}. Finally, there were various attempts and discussions  within the scope of collapse model interpretation independently from several groups \cite{Canate:2012nwv, Martin:2012pea, Das:2013qwa, Bengochea:2017cjh}. I shall only discuss the main ideas and developments based on the Collapse Model approaches in this note. 

Delicate issues related with the application of quantum theory in the cosmological framework has been stated before \cite{bell,Perez:2005gh,dan2}. There are several works where authors have studied  the emergence of classical density perturbations \cite{stan1}-\cite{stan4} which are essentially identical or very similar to the method outlined in Weinberg's monograph \cite{wein}, and which suffer from the problems I discussed in the preceding section. The emergence of classicality or dynamically breaking the symmetry of an initial quantum state without measurement is something cannot be addressed within the Copenhagen interpretation where an observer performs the act of ``measurement'' making the wavefunction collapse and only after this an outcome is achieved. As Roger Penrose had pointed there are two completely distinct processes of evolution in the Copenhagen version of quantum theory -- (i) the ``unitary'' or $U-$process which is unitary and dictates the evolution of a quantum system in isolation, and (ii) the ``reduction'' or the $R-$process dictates the reduction of quantum superposition to a stochastically chosen outcome, following the Born probability rule, once the system is measured. These two dynamical processes (i.e., unitary vs stochastic) are separated by this vague act of measurement where neither the apparatus, nor the observer or the details of measurement process are included in the theory. Although one may get around these complications in the laboratory by subjecting quantum theory as a prescription to compare the outcome of measurements, a logical extension of this outside laboratory setting is challenging. The absence of any ``observer'' or ``apparatus'' in the cosmological framework makes the quantum mechanical ``Measurement Problem'' \cite{bell-2,bell-3,tim} more explicit than in the laboratory situations \cite{bell,Perez:2005gh,dan2}.

Collapse Models \cite{col-rev-1}-\cite{Bassi:2012bg} provide a mathematically rigorous treatment to unify the $U-$process and $R$ in a single evolution equation and include the effect of apparatus in the mathematical formulation of the quantum theory. The most evolved version of Collapse Models is  called Continuous-Spontaneous-Localization (CSL) theory \cite{csl2}, {\footnote{Note that this is an evolved form of the previous discrete versions \cite{Pearle:1976ka,Ghirardi:1985mt}.}}. To understand the CSL theory we need to know  following two main equations \cite{csl2}: first, (i) a modified  quantum  dynamical   evolution, accompanied by the choice of a  certain observable $\hat A$, which is a  (stochastically) modified Schr\"odinger equation, and whose  solution is given by
  \begin{equation}\label{csl1}
 { |\psi,t\rangle_w = \hat {\cal T}e^{-\int_{0}^{t}dt'\big[i\hat H+\frac{1}{4\lambda} [w(t')-2\lambda \hat A]^{2}\big]}|\psi,0\rangle,}
\end{equation}
 where $\hat{\cal T}$ is a time-ordering operator and  $w(t)$ is a random, white noise type classical function of time. The second equation provides the probability of $w(t)$ via (ii) the Probability Distribution Density $[P(Dw(t))]$ function:
  \begin{equation}\label{csl2}
	 { P(Dw(t))\equiv{}_w \langle\psi,t|\psi,t\rangle_w \prod_{t_{i}=0}^{t}\frac{dw(t_{i})}{\sqrt{ 2\pi\lambda/dt}}}.
\end{equation}
In this way the  standard   Schr\"odinger    evolution   and the changes in the state corresponding  to a ``measurement" of  the observable $\hat{A}$ are  unified. For  non-relativistic   quantum  mechanics   of a single particle, in all situations (without invoking any  measurement device  or observer), this theory assumes a    spontaneous and continuous  reduction  characterized  by   {$\hat A = \hat {\vec X}_\delta $}, where  $\hat{\vec X}_\delta$ is  a suitably  smeared  position operator.   This can be  generalized to  multi-particle  systems and thereby everything,  including, the  apparatuses can  treated  quantum mechanically.  The theory  seems  to   successfully address the  ``measurement problem''. In order to match observational evidences at the quantitative level, the  collapse  parameter  $\lambda$ must be  small enough not to be in conflict with known tests of QM  in subatomic physics range.  However,  it also ensures rapid localization of  the ``macroscopic objects" and ensuring no superpositions in them. Recently we made a bold claim that Collapse Models can precisely identify the classical-quantum boundary \cite{Torres:2022hdv}. The originally suggested  value for the collapse parameter was $\lambda\sim 10^{-16}  sec ^{-1}$ but over the decades the CSL parameter space has being tested by several experiments and for the current status of the theory, and its empirical  constraints,  we refer the reader to consult some review articles \cite{col-rev-1}-\cite{Bassi:2012bg}.

In order to address the problems associated in standard description of the quantum generation of classical density perturbations, \cite{Canate:2012nwv} Ca\~nate, Pearle and Sudarsky (CPS) used the semiclassical gravity approach, together with collapse of the state vector according to the CSL dynamics,  where the semiclassical Einstein equation is given by
\begin{equation}\label{sc}
    G_{\mu\nu} = 8\pi \langle \Psi | T_{\mu\nu} |\Psi\rangle. 
\end{equation}
In this approach, the spacetime is treated classically but all matter fields get full quantum treatment. This equation when supplemented  by the necessary collapse of the state vector \cite{Canate:2012nwv} which makes the semiclassical treatment well-defined {\footnote{Collapse mechanism is one of the few ways to bypass the criticism on semi-classical gravity by Page and Geiker \cite{page}.}}. It is in fact very appealing that Collapse Models not only useful to address the problem at hand but also cure a fundamental inconsistency in semiclassical gravity at the same time. One major difference of this approach with the standard one described in section \ref{sec3} is that here \emph{the perturbation of the background metric does not get a quantum treatment}, just the perturbation of the matter field is quantized. In their calculations CPS chose both the field modes $\hat{\delta\phi}_k$ and corresponding momentum conjugate $\hat{\pi}_k$ as collapse generating operators with a choice of collapse parameter not just a constant but a function depending on the momentum of the field modes, chosen to be $\tilde{\lambda} = \lambda /k$ and $\tilde{\lambda} = \lambda k$, respectively. With these choices CPS made a detailed calculation, based on a CSL evolution of the state vector, in which the initial state is  the Bunch-Davies vacuum defined in the beginning of inflation, to a stochastically chosen particle excited state with some momentum $k$, as a final state after inflation, and reproduced the necessary form of the Harrison-Zel'dovich spectrum \eqref{hzs}.

In \cite{Martin:2012pea} Martin, Vennin and Peter (MVP) made another detailed study of generation of primordial perturbations using Collapse Models. They also used the CSL theory, however, chose the collapse to take place on the eigenstate of the Mukhanov-Sasaki operator $\hat{\cal{R}}_k$ rather than the field operator or the field momentum operator. Note that Mukhanov-Sasaki operator is a combination of both the perturbation of the inflaton and perturbation of the background metric. Therefore, strictly speaking MVP approach is not strictly semiclassical where background metric is never quantized, and in this sense this method is different than the method used by CPS in \cite{Canate:2012nwv}. In their study MVP found two branches of solutions in their analysis, one which reproduces the scale invariant Harrison-Zel'dovich power spectrum \eqref{hzs} and the other which does not.  The requirement that the non-scale-invariant part is outside the horizon puts some bounds on the CSL parameters which control the deviation from standard quantum prediction and MVP concluded that in the absence of any amplification mechanism the standard CSL mechanism is not strong enough to reproduce the known power spectrum in expected time frame.

Soon after the above work by MVP, Das {\it et. al.} revisited the above study \cite{Das:2013qwa} and concluded that it is in fact possible to reproduce (a) the observable power spectrum of the superhorizon modes from the appropriate branch, while (b) keeping the non scale invariant branch outside of the horizon, and (c) achieving the above two within the number of e-folds allowed by inflation (unlike MVP in \cite{Martin:2012pea}). These are possible by modifying/amplifying the effect of collapse by setting the  CSL parameter depending on the field momentum of the Mukhanov-Sasaki operator. This way of amplification of collapse mechanism is in line with what CPS used in their work \cite{Canate:2012nwv}, as well as we used for addressing the information loss during black hole evaporation process \cite{Modak:2014qja, Modak:2014vya, Bedingham:2016aus,Modak:2016uwr,Modak:2017yth}  which is another scenario where Collapse Models are very useful.

While in the above works authors always considered the Bunch-Davies vacuum state as the initial state Bengochea and Le{\'o}n performed an interesting study \cite{Bengochea:2017cjh} where they considered a more general type of state, known as the Hadamard state which in the semiclassical gravity theories provide finite values for the renormalized energy momentum tensors free from pathological behaviours such as infinite divergences. They could reproduce the observer angular power spectrum with this new choice of quantum state which puts the overall idea of using the Collapse Models on a firm ground.

More recently Martin and Vennin \cite{MV2} reported that the CMB constraints on the CSL parameter space, within their model, may invalidate the CSL theory altogether. The conclusions of their study were challenged in \cite{Bengochea:2020qsd, Bengochea:2019daa} with a further reply from the authors \cite{Martin:2020sdm}. In addition, there are more criticisms on this approach of using collapse models in the context of generation of CMB anisotropies by Kiefer and Vardanyan \cite{kief} and by Ashtekar, Corichi, and Kesavan \cite{ash} which, in turn, were also got refuted by Berjon, Okon and Sudarsky \cite{BOS}.

Recently, Lechuga and Sudarsky \cite{Lechuga:2023rvn} made an exciting proposal resolving the {\emph eternal inflation} problem which is basically a situation where the inflationary expansion of the early universe cannot be terminated just by assuming the slow-roll condition \cite{linde}. The authors show in their work that there exist a valid parameter space in the CSL based proposals which can accommodate (a) the observed anisotropies in the CMB, and (b) avoiding the necessity of eternal inflation. 

It is  worth mentioning that  in the standard  accounts of inflation, one obtains   predictions  for  the spectrum   of primordial gravity waves (i.e.  tensor modes) that are   very similar to those   for the  density perturbations,  however  the former have yet to be observed. In contrast,  the  approach described  here,    predicts primordial  gravity  waves at  a   substantially   suppressed    amplitude \cite{GW}.

All of the above works, debates, criticisms and counter criticisms make the utilization of the Collapse Models in the cosmological framework a fascinating field of study with a potential of contributing not only to cosmology but also to the foundations of the quantum theory.

\section{Building an anisotropic quantum state for the early universe}\label{sec6}
One novel feature of quantum field theory (QFT) in a curved background is the possible existence of multiple non-unitarily related  quantum vacuum states that are natural state of sets of observers who are non-inertially connected to each other. Even in the Minkowski spacetime if one is to construct a QFT in a non-inertial frame, such as for an observer with constant four momentum, we find ourselves in a similar situation where the standard vacuum in Minkowski (inertial frame) is envisioned as a particle excited state producing particles with blackbody spectrum, which is detected in the accelerating frame. This phenomena is known as Unruh effect \cite{unruh}  and it serves as a cornerstone to understand QFT in a curved space where the background spacetime is other than Minkowski. Other famous examples of particle creation, especially in curved background, are  given by the Hawking effect \cite{hawk1, hawk2} where particles are produced and black holes evaporate itself and in the  cosmological models \cite{park1,park2,park3}. 

In this section I will give a brief account of my recent works \cite{Modak:2018crw, Modak:2019jbg, Modak:2018usa, Salazar:2021pfm, Astilla:2023fph} which is a complementary line of thought to understand the early universe from a quantum perspective. We believe these studies, in future, will be able to integrate with the efforts mentioned in the last section and may provide a wholesome picture of the generation of classical density perturbations that we observe in the CMB.

The main idea is the following: if one takes semiclassical picture to generate inhomogeneity and anisotropies in the spacetime metric appearing on the left hand side of the  semiclassical Einstein equation \eqref{sc}, on the right hand side one must have a quantum state $|\Psi\rangle$ which manifestly breaks the homogeneity and isotropy symmetries. The usual Bunch-Davies vacuum state being invariant under the de Sitter symmetry group it cannot, by simple argument of isometry, be the state $|\Psi\rangle$ at the time of generation of these tiny deviations. One must find another state which is manifestly in-homogeneous (in-H) and an-isotropic (an-I). Of course, there will be  numerous conditions applicable for such a state to be well defined. Foremost, it must be a Hadamard state so that the right hand side of \eqref{sc} is free of divergences after one renormalize the stress energy tensor. Further, the resultant renormalized value of the stress tensor must satisfy the Wald axioms \cite{wald}. One may come up with several examples of such a quantum state and we can only hope that ideally one or at most a class of them could provide correct renormalized values which, via back-reaction on the metric, will   generate observed CMB inhomogeneities anisotropies. The main problem, as it stands today, is the fact that there is no such in-H and an-I state was ever built in quantum cosmology which satisfies all of the restrictions required for the well-definedness of semiclassical gravity. We find this to be a glaring omission and have made considerable progress in building one such quantum state \cite{Modak:2018crw,Modak:2018usa} and discussed various new results in this respect. Below we provide a quick review on the prescription of building the $T-$vacuum state, as termed in \cite{Modak:2018usa}, which is an in-H and an-I quantum state and lives in the radiation dominated era of the early universe and therefore it is free from inflationary effect which tends to erase any primordial inhomogeneity and anisotropy.

We shall divide our findings in two separate branches -- first (i) the geometric part, where we express the H \& I form of the radiation dominated metric into a spherically symmetric, in-H \& an-I form, and related geometric results, and second (ii) the field theoretic part, where we discuss the formulation of the $T-$vacuum and related field theoretic results. 

\subsection{Novel geometric features of radiation dominated Epoch}
Consider the spatially flat FRW metric in comoving coordinates, 
\begin{equation}\label{frw}
    ds^2 = dt^2 - a^2(t)[dr^2 + r^2(d\theta^2 + \sin^2\theta d\phi^2)],
\end{equation}
 which in cosmological time frame is
\begin{equation}\label{cfrw}
    ds^2 = a^2(\eta)[d\eta^2 - dr^2 + r^2(d\theta^2 + \sin^2\theta d\phi^2)].
\end{equation}
where the ``cosmological time'' $\eta = \int \frac{dt}{a(t)}$. Here the scale factor $a(t)$ has a dimension of length and $\eta$ is dimensionless, while $t$ has dimension of time. The only thing carry dimension \eqref{cfrw} is the conformal factor.

 Using the lightcone coordinates $u = \eta - r$ and $v = \eta + r$, (\ref{cfrw}) becomes
\begin{equation}
    ds^2 = a^2(u,v)\Big[dudv - \frac{(v-u)^2}{4}(d\theta^2 + \sin^2\theta d\phi^2)\Big].
\end{equation}
At this point make the following conformal transformation for $a(t) \propto t^{1/2}$, given by \cite{Modak:2018crw}
\begin{equation}
    U \equiv T-R = \pm \frac{\mathcal{H}e}{2}u^2, \ \ \ V \equiv T+R = \frac{\mathcal{H}e}{2}v^2,
\end{equation}
where $+$ and $-$ mean $u\geq0$ and $u\leq0$, respectively. It is straightforward to show that the above transformation implies that the radiation dominated epoch is given by
\begin{equation}\label{nm1}
    ds^2 = F_I(T, R)(dT^2 - dR^2) - R^2d\Omega^2
\end{equation}
for $U \geq 0$ ($T\geq R$), and
\begin{equation}\label{nm2}
    ds^2 = F_{II}(T,R)(dT^2 - dR^2) - T^2d\Omega^2, 
\end{equation}
for $U\leq 0$ ($T\leq R$) and the functions $F_I$ and $F_{II}$ are $F_{I}(T,R) = \frac{(\sqrt{T+R} + \sqrt{T-R})^2}{4\sqrt{T^2-R^2}}$ and $F_{II}(T,R) = \frac{(\sqrt{R+T} - \sqrt{R-T})^2}{4\sqrt{R^2-T^2}}$. These new coordinates in $(T,R)$ and $(\eta,r)$ frames are related by 
\begin{equation}\label{ts1}
    \begin{aligned}
    T &= \frac{V+U}{2} = \frac{\mathcal{H}e}{2}(\eta^2 + r^2) \\
    R &= \frac{V-U}{2} = \mathcal{H}e\eta r
    \end{aligned} 
\end{equation}
\\
for region $I$, and 
\begin{equation}\label{ts2}
    \begin{aligned}
    T &= \frac{V+U}{2} = \mathcal{H}e\eta r \\
    R &= \frac{V-U}{2} = \frac{\mathcal{H}e}{2}(\eta^2 + r^2)
    \end{aligned} 
\end{equation}
for region $II$. The conformal factors $F_{I}(T,R)$ and $F_{II}(T,R)$ as functions of the Hubble parameter for radiation stage $H = (\frac{\Dot{a}}{a})_{RD}$ in the following form \cite{Modak:2018crw} $F_{I}(H,R) = \frac{1}{1-H^2R^2}$ and $F_{II}(H,T) = \frac{1}{H^2T^2-1}$. The line $T = R$ for these observers is the comoving Hubble radius at  $R=1/H$. These metrics \eqref{nm1} and \eqref{nm2} are static up to a dynamical conformal factor and they exhibit a spherical symmetry. It was named as the $(T,R)$ frame \cite{Modak:2018usa}. 

There are few important results derived in the earlier works. For example, the $(T,R)$ frame was backtracked in the inflationary de Sitter epoch and was identified with the de Sitter static patch with a time redefinition \cite{Salazar:2021pfm}. A complete spacetime foliation with Cauchy slices was performed ensuring the well defined initial value problem \cite{Modak:2018usa} etc.

\subsection{Novel field theoretic features of radiation dominated epoch}
The motivation of constructing a field theory in the $(T,R)$ frame is the following: first we notice that the metrics in the $(T,R)$ frame as appear in \eqref{nm1} and \eqref{nm2} are inhomogeneous and anisotropic. The spherically symmetry ensures that we should be able to separate the field equation (such as the KG equation) into the radial and angular  parts. The radial part although will be dependent on coordinates $T,R$, the angular part will provide us some known solution. Given that we are able to decompose the field operator in the positive and negative frequency parts we will be able to define a new vacuum state which will be annihilated by the new annihilation operator multiplying the field modes. This new vacuum state will then also be inhomogeneous and anisotropic and if we can prove the well-definedness of this state we shall have one physically motivated, rigorously constructed anisotropic quantum state in the radiation dominated universe.

In our recent works \cite{Modak:2018usa, Astilla:2023fph} we showed that it is possible to quantize a free massless scalar field which can be related with the fluctuation of the inflaton field and with no quantization of the background metric fluctuations as necessary in a semiclassical set up. In this process we defined a manifestly in-H and an-I quantum vacuum which we called as the $T-$vacuum \cite{Modak:2018usa}. It was shown that the $T-$vacuum creates particles in the frame of cosmological and comoving observers and the distribution of particles is anisotropic in 3+1 dimensions \cite{Astilla:2023fph}. This is an exciting result since we could see the usefulness of engineering a quantum state that breaks the H \& I  symmetries. Further, we showed the Hadamard behavior in (1+1) dimensions  and a point by point comparison with the Unruh effect \cite{Modak:2018usa}. These are definitely encouraging but there remain many more unanswered questions which we want to study in future.

We are now in a position to ask the following questions, (a) could the $T-$vacuum be the yet unknown in-H and an-I state that the universe finds itself after the inflation?, (b) could we reproduce the scale invariant Harrison-Zel'dovich spectrum \eqref{hzs} by using this state?, (c) could we connect it to the studies using Collapse Models as the final state after collapse?, and finally (d) even if these ambitions are refuted could we adjust and build new states in order to match the observation in CMB? To summarize, what we are essentially asking  through all these questions is: {\emph{what is the precise quantum state of the early universe at the time of the production of density perturbations at the LSS?}}

\section{Summary and Conclusion}

To summarize, this article provides an up to date information about some efforts related with betting understanding the quantum description of the classical density perturbations from inflationary cosmology which are famously called the ``seeds of cosmic structures'' that exist today. We provided some introductory information on the detection of anisotropies in the CMB in the introduction and then discuss the observational quantity given by the Harrison-Zel'dovich power spectrum that is one of the results needs to be reproduced by any valid theory describing the problem at the hand. In this respect, we provide a brief but self contained explanation of the standard inflationary approach of using cosmological perturbations. We critically reviewed various assumption behind the calculations and discuss some of the flaws in details. Then we reviewed strong arguments present in the literature to connect the problem of quantum generation of classical density fluctuations with the foundational problems in quantum theory which cannot be bypassed in the cosmological setting. To this aspect, we reviewed recent works within the Collapse Model interpretations of quantum theory which is one of the most promising paths to overcome various flaws with the standard picture. Finally, we provided brief account of our work on building an anisotropic and physical quantum state in the post-inflationary universe, especially in the radiation dominated era, in order to look for  the precise quantum state (if any) of the early universe at the time of the production of density perturbations at the LSS.

\section*{Acknowledgements}
I thank Daniel Sudarsky and Rosa Laura Lechuga for discussions and Dhamar Astilla for reading the manuscript and correcting typos. This research is supported by CONAHCyT Grant CF/140630, Mexico. I acknowledge the Sabbatical Fellowship Program by CONAHCyT for financial support. I also acknowledge ICN-UNAM for hosting my sabbatical stay and providing necessary support for my research.

\appendix


\begin{thebibliography}{99}

\bibitem{gam}
G. Gamow, Phys. Rev. \textbf{70}, 572 (1946).

\bibitem{pen}
A. A. Penzias and R. W. Wilson, Astrophys. J. \textbf{142}, 419 (1965).



\bibitem{wein} Steven Weinberg, ``Cosmology,'' Oxford University Press (2008).

\bibitem{SZ}
Ya. B. Zel'dovich, R. A. Sunyaev, Astrophysics Space Science \textbf{4}, 301 (1969); Comments Astrophysics and Space Science \textbf{2}, 66 (1970); \textbf{4}, 173 (1972).

\bibitem{reese} E. D. Reese {\emph et al.}, Astrophysics Journal {\textbf 533}, 38 (2000).

\bibitem{SW} R. K.Sachs and A. M. Wolfe, Astrophysics Journal, {\textbf{147}}, 73 (1967).

\bibitem{HZ} E. R. Harrison, Phys. Rev, \textbf{D1}, 2726 (1970); Ya. B. Zel'dovich, MNRAS \textbf{160} 1P (1972).

\bibitem{lyth} D. H. Lyth and D. Seery, astro-ph/0607647

\bibitem{BD} T. S. Bunch and P.C.W. Davies, Proc. Roy. Society Ser. A \textbf{360}, 117 (1978).

\bibitem{MS} V. S. Mukhanov, JEPT Lett. \textbf{41}, 493 (1986), S. Sasaki, Prog. Theor. Phys., \textbf{76}, 1036 (1986).

\bibitem{Singh:2013bsa}
S.~Singh, S.~K.~Modak and T.~Padmanabhan,
 ``Evolution of quantum field, particle content and classicality in the three stage universe,''
Phys. Rev. D \textbf{88}, no.12, 125020 (2013).
\bibitem{Mahajan:2007qc}
G.~Mahajan and T.~Padmanabhan,
 ``Particle creation, classicality and related issues in quantum field theory: I. Formalism and toy models,''
Gen. Rel. Grav. \textbf{40}, 661-708 (2008).

\bibitem{Mahajan:2007qg}
G.~Mahajan and T.~Padmanabhan,
 ``Particle creation, classicality and related issues in quantum field theory: II. Examples from field theory,''
Gen. Rel. Grav. \textbf{40}, 709-747 (2008).



\bibitem{col-rev-1}A. Bassi and G. Ghirardi. Dynamical reduction models. Physics Reports, 379(5):257– 426, 2003.

\bibitem{Bassi:2023hvn}
A.~Bassi, M.~Dorato and H.~Ulbricht, ``Collapse Models: A Theoretical, Experimental and Philosophical Review,''
Entropy \textbf{25}, no.4, 645 (2023)

\bibitem{Bassi:2012bg}
A.~Bassi, K.~Lochan, S.~Satin, T.~P.~Singh and H.~Ulbricht,
 ``Models of Wave-function Collapse, Underlying Theories, and Experimental Tests,'' Rev. Mod. Phys. \textbf{85}, 471-527 (2013).

\bibitem{bell}J. Bell and A. Aspect, ``Quantum mechanics for cosmologists,'' \emph{In Speakable and Unspeakable in Quantum Mechanics: Collected Papers on Quantum Philosophy,} pages 117–138. Cambridge University Press, Cambridge, 2004.

\bibitem{Perez:2005gh}
A.~Perez, H.~Sahlmann and D.~Sudarsky,
 ``On the quantum origin of the seeds of cosmic structure,''
Class. Quant. Grav. \textbf{23}, 2317-2354 (2006), [arXiv:gr-qc/0508100 [gr-qc]].

\bibitem{dan2}D. Sudarsky, ``Shortcomings in the Understanding of Why Cosmological Perturbations
Look Classical,'' Int. J. Mod. Phys. D, 20:509–552, 2011.

\bibitem{stan1}
A. Albrecht, P. Ferreira, M.Joyce and  T. Prokopec, ``Inflation and squeezed quantum states,'' Phys. Rev. D 50 (1994) 4807.

\bibitem{stan2} D. Polarski and A. A. Starobinsky, ``Semiclassicality and decoherence of cosmological perturbations,'' Classical and Quantum Gravity 13 (1996) 377.

\bibitem{stan3} F. C. Lombardo and D. Lopez Nacir, ``Decoherence during inflation: The generation of classical inhomogeneities,'' Phys. Rev. D 72, 063506, (2005)

\bibitem{stan4} C. Kiefer, D. Polarski, ``Why do cosmological perturbations look classical to us?'', Adv.Sci.Lett.2:164-173,2009 [arXiv: astro-ph/0810.0087].


\bibitem{bell-2}John Bell, ``Against Measurement,''  Phys. World 3 (8) 33 (1990).

\bibitem{bell-3} J.S. Bell, ``Speakable and unspeakable in Quantum Mechanics: Collected Papers on Quantum Philosophy,'' Cambridge University Press, Cambridge, (1987).

\bibitem{tim} T. Maudlin, ``Three measurement problems,'' Topoi, 14(1), 7–15, (1995).

\bibitem{mwi} J. Hartle and T. Hertog, ``One Bubble to Rule Them All,'' Phys. Rev. D, 95(12):123502, 2017; E. Okon and D. Sudarsky, ``On the Consistency of the Consistent Histories Approach to Quantum Mechanics,'' Found. Phys., 44:19–33, 2014.

\bibitem{BM}Roderich Tumulka, ``Bohmian Mechanics,'' Chapter 17: The Routledge Companion to Philosophy of Physics, Routledge (2021), [arXiv:1704.08017 [quant-ph]].




\bibitem{Goldstein:2015mha}
S.~Goldstein, W.~Struyve and R.~Tumulka,
 ``The Bohmian Approach to the Problems of Cosmological Quantum Fluctuations,''
[arXiv:1508.01017 [gr-qc]].

\bibitem{Pinto-Neto:2018zvn}
N.~Pinto-Neto and W.~Struyve,
 ``Bohmian quantum gravity and cosmology,''
[arXiv:1801.03353 [gr-qc]].

\bibitem{Torres:2022hdv}
F.~Torres, S.~K.~Modak and A.~Aranda,
 ``New Insights on the Quantum-Classical Division in Light of Collapse Models,''
Found. Phys. \textbf{53}, no.4, 73 (2023)

\bibitem{Canate:2012nwv}
P.~Ca\~nate, P.~Pearle and D.~Sudarsky,
 ``Continuous spontaneous localization wave function collapse model as a mechanism for the emergence of cosmological asymmetries in inflation,''
Phys. Rev. D \textbf{87}, no.10, 104024 (2013).

\bibitem{Martin:2012pea}
J.~Martin, V.~Vennin and P.~Peter,
 ``Cosmological Inflation and the Quantum Measurement Problem,''
Phys. Rev. D \textbf{86}, 103524 (2012)

\bibitem{Das:2013qwa}
S.~Das, K.~Lochan, S.~Sahu and T.~P.~Singh,
 ``Quantum to classical transition of inflationary perturbations: Continuous spontaneous localization as a possible mechanism,''
Phys. Rev. D \textbf{88}, no.8, 085020 (2013)
[erratum: Phys. Rev. D \textbf{89}, no.10, 109902 (2014)]

\bibitem{Modak:2014qja}
S.~K.~Modak, L.~Ort\'\i{}z, I.~Pe\~na and D.~Sudarsky,
 ``Black hole evaporation: information loss but no paradox,''
Gen. Rel. Grav. \textbf{47}, no.10, 120 (2015).

\bibitem{Modak:2014vya}
S.~K.~Modak, L.~Ort\'\i{}z, I.~Pe\~na and D.~Sudarsky,
 ``Non-Paradoxical Loss of Information in Black Hole Evaporation in a Quantum Collapse Model,''
Phys. Rev. D \textbf{91}, no.12, 124009 (2015).

\bibitem{Bedingham:2016aus}
D.~Bedingham, S.~K.~Modak and D.~Sudarsky,
 ``Relativistic collapse dynamics and black hole information loss,''
Phys. Rev. D \textbf{94}, no.4, 045009 (2016).

\bibitem{Modak:2016uwr}
S.~K.~Modak and D.~Sudarsky,
``Modelling non-paradoxical loss of information in black hole evaporation,''
Fundam. Theor. Phys. \textbf{187}, 303-316 (2017).

\bibitem{Modak:2017yth}
S.~K.~Modak and D.~Sudarsky,
 ``Collapse of the wavefunction, the information paradox and backreaction,''
Eur. Phys. J. C \textbf{78}, no.7, 556 (2018)




\bibitem{Bengochea:2017cjh}
G.~R.~Bengochea and G.~Le\'on,
 ``Novel vacuum conditions in inflationary collapse models,''
Phys. Lett. B \textbf{774}, 338-350 (2017)

\bibitem{Martin:2021lje}
J.~Martin and V.~Vennin,
 ``On the choice of the collapse operator in cosmological Continuous Spontaneous Localisation (CSL) theories,''
Eur. Phys. J. C \textbf{81}, no.6, 516 (2021).


\bibitem{page}D. N. Page and C. D. Geilker, ``Indirect evidence for quantum gravity,'' Phys. Rev. Lett. 47, 979 (1981).

\bibitem{MV2} J.~Martin and V.~Vennin, ``Cosmic Microwave Background Constraints Cast a Shadow On Continuous Spontaneous Localization Models,'' Phys. Rev. Lett. 124, 080402 (2020).

\bibitem{Bengochea:2020qsd}
G.~R.~Bengochea, G.~Leon, P.~Pearle and D.~Sudarsky,
 ``Comment on ''Cosmic Microwave Background Constraints Cast a Shadow On Continuous Spontaneous Localization Models'','' [arXiv:2006.05313 [gr-qc]].

\bibitem{Bengochea:2019daa}
G.~R.~Bengochea, G.~Le\'on, E.~Okon and D.~Sudarsky,
 ``Can the quantum vacuum fluctuations really solve the cosmological constant problem?,''
Eur. Phys. J. C \textbf{80}, no.1, 18 (2020).

\bibitem{Martin:2020sdm}
J.~Martin and V.~Vennin,
 ``A response to criticisms on ''CMB Constraints Cast a Shadow on CSL Model'',''
Eur. Phys. J. C \textbf{81}, no.1, 64 (2021)
 
\bibitem{kief} C. Kiefer and T. Vardanyan, ``Power spectrum for perturbations in an inflationary
model for a closed universe,'' Gen. Rel. Grav., 54(4):30, 2022

\bibitem{ash}A. Ashtekar, A. Corichi, and A. Kesavan, ``Emergence of classical behavior in the early
universe,'' Phys. Rev. D, 102(2):023512, 2020.

\bibitem{BOS}J. Berjon, E. Okon, and D. Sudarsky, ``Critical review of prevailing explanations for the
emergence of classicality in cosmology,'' Phys. Rev. D, 103(4):043521, 2021.

\bibitem{Lechuga:2023rvn}
R.~L.~Lechuga and D.~Sudarsky,
 ``Eternal inflation and collapse theories,''
JCAP \textbf{01}, 038 (2024).

\bibitem{linde}A. Linde, ``Eternal inflation,'' in ``Inflationary cosmology'', Springer 2010 [Editors: Martin Lemoine, Jerome Martin, Patrick Peter].

\bibitem{GW} 
G. León, L. Kraiselburd, and S. J. Landau, ``Primordial gravitational waves and the collapse of the wave function,''  Phys. Rev. D 92, 083516 (2015).\\
T. Markkanen, S. Rasanen, and P. Wahlman, ``Inflation without quantum gravity,''  Phys. Rev. D 91,  084064 (2015).\\
G. León, A. Majhi, E. Ok\'on, and D. Sudarsky, ``Reassessing the link between B-modes and inflation,'' Phys. Rev. D 96, 101301(R) (2017).\\
G. León, A. Majhi, E. Ok\'on, and D. Sudarsky, ``Expectation of primordial gravity waves generated during inflation,'' Phys. Rev. D 98,  023512 (2018). 



\bibitem{csl2}G. C. Ghirardi, P. Pearle and A. Rimini, ``Markov processes in Hilbert space and continuous spontaneous localization of systems of identical particles,'' Phys. Rev. A42, 78 (1990).


\bibitem{unruh}
W. G. Unruh, Notes on black-hole evaporation, Phys. Rev. D 14, 870 (1976).



\bibitem{hawk1}
S. W. Hawking, Black hole explosions?, Nature 248, 30 (1974).

\bibitem{hawk2}
S. W. Hawking, Particle creation by black holes, Commun. Math. Phys. 43 (1975) 199.

\bibitem{park1}
L. Parker, Particle creation in expanding universes, Phys. Rev. Lett. 21, 562 (1968).

\bibitem{park2}
L. Parker, Phys. Rev. 183, 1057 (1969).

\bibitem{park3}
L. Parker, Quantized Fields and Particle Creation in Expanding Universes. II, Phys. Rev. D 3, 346 (1971) Erratum: [Phys.Rev. D 3, 2546 (1971)].





\bibitem{Modak:2018crw}
S.~K.~Modak,
 ``New geometric and field theoretic aspects of a radiation dominated universe,''
Phys. Rev. D \textbf{97} (2018) no.10, 105016

\bibitem{Modak:2019jbg}
S.~K.~Modak,
 ``Cosmological Particle Creation Beyond de Sitter,''
Int. J. Mod. Phys. D \textbf{28} (2019) no.09, 1930015

\bibitem{Modak:2018usa}
S.~K.~Modak,
 ``Physical observers, $T$-vacuum and Unruh like effect in the radiation dominated early universe,''
JHEP \textbf{12}, 031 (2020).

\bibitem{Salazar:2021pfm}
J.~R.~Salazar and S.~K.~Modak,
 ``Quantum field theory in a de-Sitter universe transiting to the radiation stage,''
JHEP \textbf{05}, 048 (2022)

\bibitem{Astilla:2023fph}
D.~S.~Astilla, S.~K.~Modak and E.~Salazar,
 ``Anisotropic particle creation from $T-$vacuum in the radiation dominated universe,''
[arXiv:2312.17129 [gr-qc]].



\bibitem{wald}
R. Wald, , ``General Relativity'', University of Chicago Press (1984).

\bibitem{Pearle:1976ka}
P.~M.~Pearle,
 ``Reduction of the State Vector by a Nonlinear Schrodinger Equation,''
Phys. Rev. D \textbf{13}, 857-868 (1976).

\bibitem{Ghirardi:1985mt}
G.~C.~Ghirardi, A.~Rimini and T.~Weber,
 ``A Unified Dynamics for Micro and MACRO Systems,''
Phys. Rev. D \textbf{34}, 470 (1986).




\end{thebibliography}
\end{document}